\documentclass[conference]{IEEEtran}
\IEEEoverridecommandlockouts
\usepackage{cite}
\usepackage{graphicx}
\usepackage{textcomp}
\usepackage{xcolor}
\def\BibTeX{{\rm B\kern-.05em{\sc i\kern-.025em b}\kern-.08em
    T\kern-.1667em\lower.7ex\hbox{E}\kern-.125emX}}

\usepackage{enumitem}
\setlist{nosep}
\usepackage[font=small,skip=6pt]{caption}
\usepackage{subcaption}
\usepackage{tikz}
\usetikzlibrary{positioning}

\usepackage[group-separator={,}]{siunitx}

\usepackage{amsmath,amsfonts,amssymb,amsthm}
\usepackage{mathtools}
\usepackage{commath}
\newtheorem{theorem}{Theorem}[section]
\newtheorem{definition}{Definition}[section]

\usepackage{algorithm}
\usepackage[noend]{algpseudocode}
\usepackage{tabularx}
\usepackage{multicol}
\usepackage{multirow}
\usepackage{booktabs}

\makeatletter
\algrenewcommand\ALG@beginalgorithmic{\small}
\makeatother

\begin{document}

\title{A Differentially Private Algorithm for Range Queries on Trajectories}

\author{\IEEEauthorblockN{Soheila Ghane, Lars Kulik, Kotagiri Ramamohanarao}
\IEEEauthorblockA{\textit{Department of Computing and Information Systems} \\
\textit{The University of Melbourne, Australia}\\
Email: soheila.ghane@student.unimelb.edu.au, \{lkulik, kotagiri\}@unimelb.edu.au}
}

\maketitle

\begin{abstract}
We propose a novel algorithm to ensure $\epsilon$-differential privacy for answering range queries on trajectory data. In order to guarantee privacy, differential privacy mechanisms add noise to either data or query, thus introducing errors to queries made and potentially decreasing the utility of information. In contrast to the state-of-the-art, our method achieves significantly lower error as it is the first data- and query-aware approach for such queries. The key challenge for answering range queries on trajectory data privately is to ensure an accurate count. Simply representing a trajectory as a set instead of \emph{sequence} of points will generally lead to highly inaccurate query answers as it ignores the sequential dependency of location points in trajectories, i.e., will violate the consistency of trajectory data. Furthermore, trajectories are generally unevenly distributed across a city and adding noise uniformly will generally lead to a poor utility. To achieve differential privacy, our algorithm adaptively adds noise to the input data according to the given query set. It first privately partitions the data space into uniform regions and computes the traffic density of each region. The regions and their densities, in addition to the given query set, are then used to estimate the distribution of trajectories over the queried space, which ensures high accuracy for the given query set. We show the accuracy and efficiency of our algorithm using extensive empirical evaluations on real and synthetic data sets.
\end{abstract}

\begin{IEEEkeywords}
Spatial histogram, Trajectory, Range query, Differential privacy
\end{IEEEkeywords}

\section{Introduction}
\begin{figure*}[htb]
\centering
\includegraphics[width=0.8\linewidth]{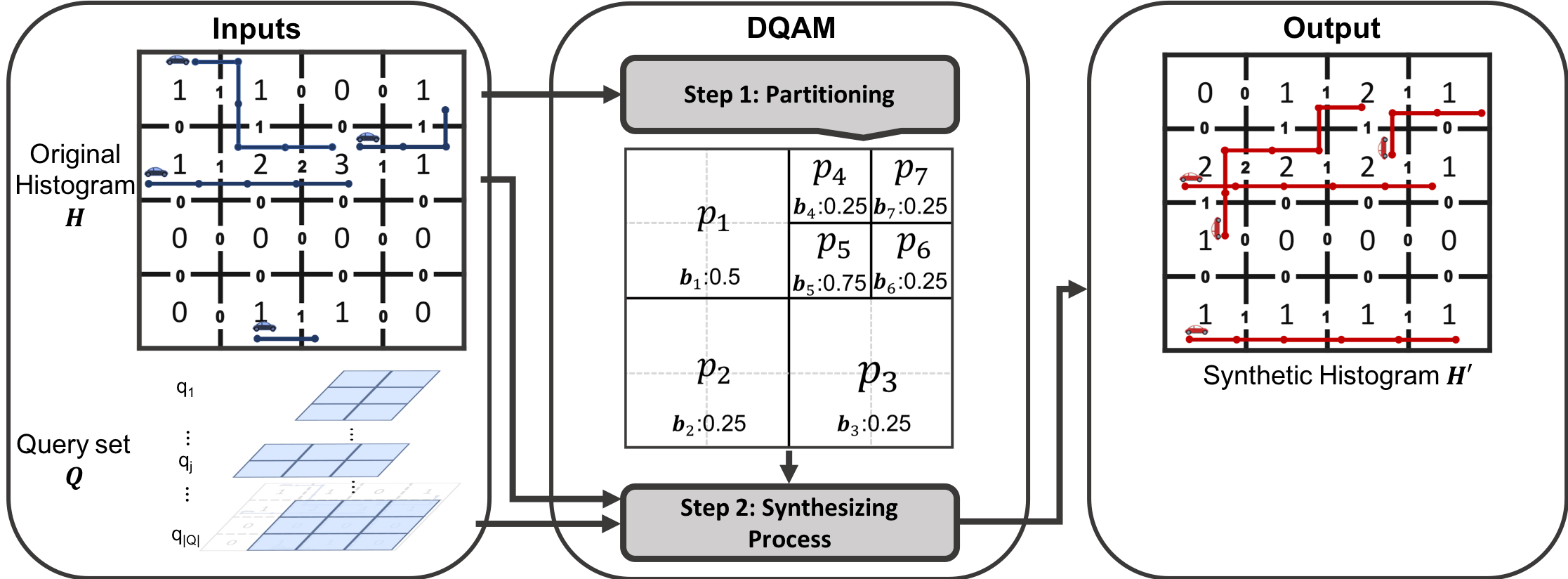}
\caption{Overview of \textit{DQAM} mechanism with an example. $p_{i}$ and $b_{i}$ depict the partition and its density, respectively.}
\label{fig:overview}
\end{figure*}

The popularity of sensor-enabled devices (e.g., wearables and smartphones) has significantly advanced the capability of businesses to collect and analyze people's trajectories. Studying large-scale data sets and analyzing the movement patterns of individuals provide crucial insight for many applications (e.g., traffic management, route planning, urban planning, crime detection). Such applications critically rely on estimating the number of trajectories in an area. For example, in urban planning, computing the number of pedestrians and the flow of their movements provide significant information for the placement of public spaces (e.g., local parks, street spaces, plazas), pedestrian and bicycle paths, and public transport stations. A fundamental query type in studying the movement patterns is \emph{range query} on trajectories~\cite{leonardi2014general, lopez2005spatiotemporal} which counts the number of \emph{distinct} trajectories intersecting the query area (2D space). However, computing the cost of this query type on raw trajectories is linear in the number of trajectories and the average length of the trajectories. To avoid such high computational cost \emph{spatial histogram} (see Section~\ref{sec:Background}) is commonly used to compute range queries efficiently. The cost of range queries using spatial histogram is linear in the size of the query area. A spatial histogram considers the data space as a grid with fixed cell size and aggregates the count of trajectories overlapping a cell or an incident edge of two cells in the grid. In other words, a spatial histogram maps the trajectory data into a set of counts representing the distribution of trajectories throughout the space.

Even though the value of spatial histograms in studying movement patterns is realized by companies, releasing such data to third parties is still a significant issue due to privacy concerns. Studies show that trajectories exhibit a high risk of being identified that aggregation or coarsening techniques as done in a spatial histogram, barely reduce their uniqueness~\cite{xu2017trajectory, de2013unique}. The key reason is that trajectories are time-stamped \emph{sequences} of locations with a start and end points. For example, the start of a trajectory is often a home location. Thus, publishing a spatial histogram raises privacy concerns due to the uniqueness of people’s trajectories, particularly in sparse regions with few trajectory counts. 

A principled approach is needed to generate a synthetic spatial histogram with strong privacy guarantees of differential privacy~\cite{Dwork2006} while ensuring high utility in practice. The utility refers to the accuracy in answering range queries. The approach can make use of a set of range queries to learn the distribution of data (called \emph{query-awareness}). However, it should ensure the synthesized spatial histogram maintains high accuracy in answering any range queries on trajectories. A naive approach adds an equal noise to the cell counts in the histogram to ensure privacy. Since a trajectory may intersect multiple cells, the cells are \textit{sequentially dependent}. Thus, naively scaling noise to satisfy differential privacy can violate the sequentiality between the cells, which is crucial in representing trajectories. Furthermore, the sensitivity of cell counts to the added noise varies depending on the \emph{density} of trajectories in different regions of data space. A large region of space might be sparse with small cell counts which makes the cell counts in the region highly sensitive to the added noise. In contrast, a region might be small but very dense such that cell counts are large and highly noise resistance. The proposed approach should be able to ensure maintaining the data properties (called \emph{data-awareness}) in the synthesized data.

In this paper, we present \textit{Data- and Query-Aware Mechanism (\textit{DQAM})}, a mechanism that synthesizes spatial histograms. As~\cite{Ghane2018Publishing}, \textit{DQAM} takes a spatial histogram and a set of range queries as input and uses the correlation between the queries (i.e., is \textit{query-aware}) to synthesize the histogram. In contrast to~\cite{Ghane2018Publishing}, \textit{DQAM} identifies the density of different regions in data and make use of the density in scaling the added noise to the generated synthetic histogram (i.e., is \emph{query-aware}). Also, \textit{DQAM} generates the optimal synthetic histogram in each iteration of the synthesizing process that satisfies the sequentiality of cells. The proposed mechanism in~\cite{Ghane2018Publishing} applies a greedy strategy to satisfy the sequentiality requirement among histogram cells, which may lead to overcorrection. While the work in~\cite{Ghane2018Publishing} successfully deals with trajectories uniformly distributed in the data space, it can result in considerable loss of utility for trajectory data sets with a non-uniform distribution in different regions. \textit{DQAM} is designed to deal with such realistic trajectory distributions by capturing the density of trajectories in different regions and using the density as a weighting measure: each region is assigned a weight relative to its density. To guarantee differential privacy, \textit{DQAM} uses the well established Laplace mechanism~\cite{Dwork2006} and Exponential mechanism~\cite{mcsherry2007} for adding noise in all parts of its computations. 

\textbf{Contributions.} First, we propose a differentially private \textit{Data- and Query-Aware Mechanism} (\textit{DQAM}), a mechanism that publishes a spatial histogram to answer range queries efficiently. To the best of our knowledge, \textit{DQAM} is the first data- and query-aware mechanism for range queries on sequential spatial data that reduces the error by a factor up to 7.4 compared to the current approaches on trajectory data sets. Second, we design an efficient data-aware algorithm with $nlog(n)$ time complexity in the number of cells in the spatial histogram that partitions the histogram into uniform regions. Third, we present a query-aware and differentially private strategy that captures trajectory sequentiality and synthesizes accurate output using the given queries and partitions.

\subsection*{Mechanism Overview} \label{sec:mechanism overview}
Differential privacy assumes a privacy budget for an algorithm that is described by a parameter $\epsilon$. \textit{DQAM} is a two-stage $\epsilon$-differentially private mechanism for answering range queries on trajectories (see Fig.~\ref{fig:overview}). The mechanism takes as input a set of range queries $Q$, and a spatial histogram $H$. The histogram $H$ is a representation of trajectory counts throughout the data space. The query set $Q$ is randomly generated and is large to cover the entire space of $H$. The output of \textit{DQAM} is $H'$, a private estimate of $H$, which preserves the distribution of original histogram. In first stage, the histogram is partitioned into \textit{disjoint} and \textit{uniform} regions and the trajectory density of each region is computed. Intuitively, partitioning the histogram space into uniform regions improves the signal to noise ratio, which in turn improves the accuracy of the estimation process in stage 2. Instead of estimating the density of every single cell, the algorithm estimates the density of a region and assumes the same density for all cells in the region. Additionally, maintaining the density of a uniform region is significantly easier than a non-uniform region. Since the trajectory data sets are usually sparse, there are potentially many regions with (semi-)uniform distribution in the data. Utilizing this property, the estimation algorithm considers a uniform region as a single entity and uses its density in the private estimation process. Furthermore, the requirement that generated partitions should be disjoint, ensures that each region can be treated independently.

Any algorithm that efficiently generates disjoint and uniform partitions can be used in this stage of \textit{DQAM}. One approach is to generate all possible partitions and compute the uniformity cost of each partition. The partitioning algorithm should find a set of these partitions, which are disjoint, cover the entire space, and have the minimum uniformity cost. However, there is no closed-form expression for computing all possible partitions in a $m \times n$ space and the computational complexity grows with the order of $O(2^{m \times n})$. Furthermore, finding the set with the minimum cost is computationally expensive. Another approach is using space partitioning techniques such as quadtree, kd-tree and R-tree. However, the R-tree and kd-tree cannot adapt well to identify disjoint and uniform regions of trajectories. An R-tree cannot cover areas with no trajectory, and the output partitions may be overlapped. A kd-tree requires an ordering among trajectories to do partitioning while neither trajectories nor their counts in the histogram have any order. In addition, the partitioning in kd-tree is based on the density of regions while the outputs might not have a uniform distribution. In contrast, a quadtree decomposes the space into disjoint regions. By defining a partitioning cost to measure the uniformity of regions, a quadtree can partition the space into uniform regions. Whilst, the generated partitions may not be optimal concerning the uniformity cost, the quadtree is computationally efficient, and the final partitioning provides an accurate estimation of uniform regions. However, the regions in a quadtree branch may overlap which could be used to refine parts of a user's trajectory. We need to consider this correlation in the variance of added noise to ensure differential privacy.

Note that regardless of the approach used in stage one, only the outputs (i.e., final partitioning) are used for the next stage of \textit{DQAM}. Thus, we need to ensure that the final partitions are differentially private as the rest of the generated partitions during the process will be removed. For example, in a constructed quadtree, we need to ensure the privacy of leaves.

In the second stage, \textit{DQAM} uses the computed final partitions and densities, in addition to the given query set $Q$, to estimate the distribution of original histogram $H$. The output is an $\epsilon$-differentially private histogram $H'$ that is generated using the data and query properties, i.e., \textit{DQAM} is data- and query-aware. Since the two stages iteratively access the original histogram, we split the $\epsilon$ budget. We will show that we need in total four privacy budgets $\epsilon_i, i \in \{1,\ldots,4\}$ with $\sum_{i=1}^4 \epsilon_i = \epsilon$. We conducted a pre-study that investigated the impact of different weightings but found no significant impact on the accuracy/utility and hence, set $\epsilon_{i} = \frac{\epsilon}{4}$.

\textbf{Step 1: Private Partitioning.} 
We develop an efficient algorithm based on the two approaches described above that defines the uniformity as a cost function and partitions the histogram into uniform regions. Intuitively, in a uniform region, all values can be manipulated/treated equally, which significantly improves the accuracy of the synthesizing process (see Section~\ref{subsec:Quadtree search}). It generates a list of uniform regions $P$ (i.e., the quadtree leaves) representing disjoint partitions of the histogram. The outputs of this stage are the partitions ($P$) and their densities ($B$), i.e., the number of trajectories in a region relative to the total number of trajectories in the entire histogram. Since the partitioning process and computing densities need interacting with the original histogram, we use the Laplace mechanism~\cite{Dwork2006} to add noise to them.

\textbf{Step 2: Private Synthesizing Process.}
In this step, our algorithm takes the partitions $P$, densities $B$ and the given range queries $Q$ to privately learn the distribution of the original spatial histogram $H$ and generate an $\epsilon$-differentially private histogram $H'$. The learning process is iterative and starts with a uniform trajectory distribution. In each iteration, one query $q \in Q$ is selected. Based on the query error on $H'$, the estimation is updated by rescaling up/down the values in $H'$. The values in a region $p \in P$ are rescaled if $p$ overlaps with $q$. The density of region $p$ magnifies/reduces the update scale for that region. Previous work on estimating histogram values while ensuring \textit{local sequentiality}~\cite{Ghane2018Publishing} used a heuristic strategy to preserve the dependency. Our paper presents an algorithm to compute the optimal estimation for $H'$ while ensuring the dependency. For the query selection and adding noise to the query answer, the exponential and Laplace mechanisms are applied, respectively.

The following example is a sample execution of \textit{DQAM}.

\textsc{\textbf{Example 1.}} \emph{An overview of DQAM is presented in Fig.~\ref{fig:overview}. The inputs are a histogram $H$ with $4$ trajectories and the set of range queries $Q$ shown graphically as rectangular areas over the histogram. A possible output of partitioning can be $P = \{p_{1}, ..., p_{7}\}$. This need not to be optimal as defined in Section~\ref{sec:Private Partitioning} because the partitioning is randomized. The density of partitions are $B = \{b_{1}=2/4=0.5, b_{2}=1/4=0.25, b_{3}=1/4=0.25, b_{4}=1/4=0.25, b_{5}=3/4=0.75, b_{6}=1/4=0.25, b_{7}=1/4=0.25\}$. The $P$ and $B$ in addition to $Q$ are then used in step 2 to generate $H'$.} 



\section{Background} \label{sec:Background}
\subsection{Spatial Histogram and Range Queries}

\begin{figure}[t!] 
    \centering
    \includegraphics[width=0.9\linewidth]{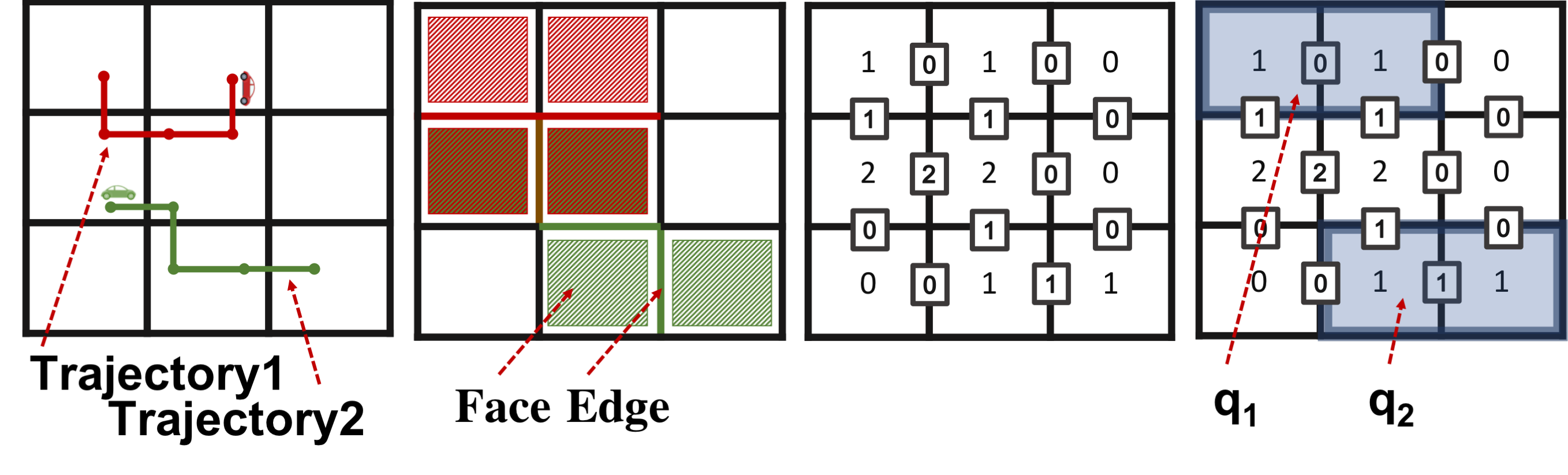}
    \caption{A data set with two trajectories and its corresponding spatial histogram; two examples of range queries ($q_{1}$ and $q_{2}$).}
    \label{fig:SpatialHist}
    
\end{figure}

Consider a range query with a convex area over a trajectory data set $D$ which counts the distinct number of trajectories intersecting the area. Let $D$ be a subset of size $n$ of the trajectory universe $\mathcal{T}$ of trajectories without loops. Each trajectory $t \in D$ has a domain equal to the data space $\mathcal{S} = \{s_{1}, s_{2}, ..., s_{N}\}$ where $s_{i} \in \mathcal{S}$ is a location given by longitude and latitude. To express the query, we first map $D$ into a grid-based histogram called \emph{spatial histogram} (see Fig.~\ref{fig:SpatialHist}). 

A spatial histogram $H=\{F, E_{v}, E_{h}\}$ with $r$ rows and $c$ columns, consist of three components: (1) faces $F \in \mathbb{N}_{0}^{r \times c}$, (2) vertical edges $E_{v} \in \mathbb{N}_{0}^{r \times (c-1)}$ and (3) horizontal edges $E_{h} \in \mathbb{N}_{0}^{(r-1) \times c}$. In particular, $H$ maps $D$ into three matrices describing the overlap of trajectories with the histogram cells, i.e., $H:D \rightarrow \mathcal{H}$, where $\mathcal{H} = \{\mathbb{N}_{0}^{r \times c}, \mathbb{N}_{0}^{r \times (c-1)}, \mathbb{N}_{0}^{(r-1) \times c}\}$ is the histogram domain. The edge components (small boxes in Fig.~\ref{fig:SpatialHist}) distinguish a spatial histogram from simple histograms by recording the count of trajectories crossing the edge between two adjacent cells. The edge components preserve the \emph{sequential dependency} of locations in a trajectory referred to as \emph{local sequentiality}. We use $E = \{E_{v}, E_{h}\}$ from now on to refer to both the edge components. To make the presentation simple, we use $e \in E$ to denote an edge $e$ in which $e \in E_{v}$ or $e \in E_{h}$. Without loss of generality we assume each histogram cell $H[j]$ in index $j = (row, col)$ is a square of fixed side length and all cells have the same size. 

Spatial histograms are known as data structures to efficiently store trajectories and compute range queries~\cite{xie2007, xie2014}. The shaded areas $q_{1}$ and $q_{2}$ in Fig.~\ref{fig:SpatialHist} show two range queries. We assume a range query specifies a rectangular region of
$r' \geq r$ cells height and $c' \geq c$ cells width, where the axes are aligned with the cells' boundaries. A range query can be answered by subtracting the sum of edge values from the sum of face values, which are both \textit{inside} the query area. For instance, the total number of trajectories in $q_{2}$ is $2 - 1 = 1$. Note the \textit{border edges are not counted} in computing the query answer because spatial histograms are constructed in such a way that trajectories can only start and end in faces.

\begin{definition}
A range query $q=\{q_{F}, q_{E} \mid q_{F} =$ set of faces inside the query rectangle q and  $q_{E} =$ set of edges inside the query rectangle $q\}$ is as a function that takes a histogram $H=\{F,E\}$ as input and maps it to a non-negative integer values, i.e., $q:\mathcal{H} \rightarrow \mathbb{N}_{0}$, by counting the distinct trajectories intersecting the query area: \[q(H) = q_{F}(F) - q_{E}(E),\]
where $q_{F}(F)=q_{F} \diamond F$ and $q_{E}(E)=q_{E} \diamond E$, and ``$\diamond$'' is the \textit{dot product} defined over matrices (i.e., the sum of element-wise multiplication of the matrices. This can also be written as trace of $(q_{F}*F^{\intercal})$).
\end{definition}

Note that $q_{F}$ and $q_{E}$ are binary matrices indicating the query area. Our definition would translate to the general definition of a range query if we used the lower and upper column and row indices. A trajectory may have multiple intersections with the query area (e.g., $q_{1}$ in Fig.~\ref{fig:SpatialHist}). Computing the range query for such cases requires modified versions of spatial histograms~\cite{xie2010, xie2014}. For the sake of simplicity, we use the basic histogram defined above, but the results can be generalized. The notations of the paper are summarized in Table~\ref{table:notation}.

\begin{table}[t!]
    \centering
    \begin{tabular}{|c|l|}
        \hline
        $D$ & Trajectory data set, $D \subseteq \mathcal{T}$ and $|D| = n$ \\
        \hline
        $H$ & Spatial histogram related to $D$, $H = \{F, E\}$, $H \subseteq \mathcal{H}$ \\
        \hline
        $E$ & The horizontal and vertical edges in $H$, $E = \{E_v, E_h\}$ \\
        \hline
        $j$ & Index of a cell in histogram, $j = (row, col)$ \\
        \hline
        $H[j]$ & A cell in $H$ with a face $f \in F$ and four edges $e \in E$ \\
        \hline
        $f_j$ & The face component of cell $H[j]$ \\
        \hline
        $e_{ij}$ & The incident edge of two adjacent cells $H[i]$ and $H[j]$ \\
        \hline
        $H'$ & Synthetic spatial histogram \\
        \hline
        $H''$ & Intermediate synthetic histogram in the synthesizing process \\
        \hline
        $T$ & Total number of iteration in the synthesizing process \\
        \hline
        $Q$ & Set of queries $q \in Q$ that $q = \{q_F, q_E\}$ \\
        \hline
        $q[j]$ & The $j$th element of query, $q[j] \in \{0, 1\}$ \\
        \hline
        $q(H)$ & Computing the query $q$ on histogram $H$ \\
        \hline
        $P$ & Set of partitions $p_i \in P$ \\
        \hline
        $p_i[j]$ & The $j$th cell in the region $p_i$ \\
        \hline
        $B$ & Set of partition densities $b_i \in B$, $b_i = \sum_j c(p_i[j])/|D|$ \\
        \hline
        $\delta$ & The threshold of uniformity cost in partitioning (Algorithm~\ref{alg:partitioning}) \\
        \hline
        $werr_i$ & Error of selected query at iteration $i$ (Algorithm~\ref{alg:data and query adaptive estimation}) \\
        \hline
        $c(x)$ & The count of $x$ \\
        \hline
        $|x|$ & Total number of elements/cells in $x$ \\
        \hline
    \end{tabular}
    \caption{Summary of notation.}
    \label{table:notation}
\end{table}

\subsection{Differential Privacy}
\label{subsec:DP}
According to differential privacy~\cite{Dwork2006}, the difference between the probability of mechanism outputs, in the presence/absence of a record in the input data set, is bounded by a privacy budget $\epsilon$. For a trajectory data set $D$, the set $nbr(D)$ denotes data sets that are different from $D$ in \textit{at most one trajectory} (by changing, adding or removing a trajectory), i.e., $nbr(D)=\{ D^*\big||D-D^*|=1\}$. We refer to the spatial histogram $H$ instead of $D$, since $H$ represents a mapping of the trajectory data set $D$ and changing a trajectory in $D$ directly affects the counts in $H$.

\begin{definition}
Given a histogram $H \subseteq \mathcal{H}$ and any histogram $H^* \in nbr(H)$, a randomized mechanism $\mathcal{M}$ is $\epsilon$-differentially private, $\epsilon > 0$, and any output $O\subseteq Range(\mathcal{M})$ if $Pr[\mathcal{M}(H) \in O] \leq exp(\epsilon) . Pr[\mathcal{M}(H^*) \in O]$.
\end{definition}

Differential privacy has two important composition properties~\cite{mcsherry2009}. Let $\mathcal{M}_{1}, \mathcal{M}_{2},... \mathcal{M}_{n}$ be $n$ mechanisms each satisfying $\epsilon_{i}$-differential privacy. When $\mathcal{M}_{1}, \mathcal{M}_{2},... \mathcal{M}_{m}$ are executed sequentially on a histogram, it is called \textit{sequential composition} and the output is $(\sum \epsilon_{i})$-differentially private. When the histogram is partitioned into $n$ disjoint subsets and $\mathcal{M}_{i}$ is applied on subset $i$, it is called \textit{parallel composition} and the output is $\{\max{\epsilon_{i}}\}$-differentially private. Since in a spatial histogram a trajectory is represented as a sequence of cells, each count can be treated separately, i.e., parallel composition. However, a naive approach may violate the local sequentiality of cells, which significantly reduces the utility of the output. Our proposed mechanism guarantees the local sequentiality in the released histogram and thus achieves high utility.

For a range query $q$ differential privacy can be achieved by adding a properly scaled random noise to the output. The noise scale depends on $q$'s \emph{sensitivity}~\cite{Dwork2006}, which captures the maximum difference in query answers between any two neighboring histograms:

\begin{definition}
The sensitivity of $q: \mathcal{H} \rightarrow \mathbb{N}_{0}$, denoted by $\Delta q$, is $\Delta q = \max\limits_{H,H^* \in nbr(H)} ||q(H)-q(H^*)||_{1}$.
\end{definition}

Since adding (removing) one trajectory to (from) a spatial histogram changes the answer of a range query at most by one, \emph{the sensitivity of a range query is $1$}. The Laplace mechanism (LM)~\cite{Dwork2006} achieves differential privacy by adding a Laplace noise to the query's answer. We use Lap($\sigma$) to denote the Laplace probability distribution with mean $0$ and scale $\sigma$.

\begin{definition}
Given a histogram $H \subseteq \mathcal{H}$ and a range query $q: \mathcal{H} \rightarrow \mathbb{N}_{0}$, the Laplace mechanism is defined as $ LM(H, q)= q(H) + Lap(\Delta q/\epsilon)^{|H|}$.
\end{definition}

We apply the Laplace mechanism in the first step of \textit{DQAM} where the cost and density of a partition are functions of a range query. The second step, however, is a non-numerical function that selects a query $q$ from $Q=\{q_{i}| i \in \{1, 2, ..., m\} \}$. To achieve differential privacy, the Exponential Mechanism (EM)~\cite{mcsherry2007} randomly selects a query with the highest utility on the estimated histogram $H'$. In the Exponential mechanism, a random noise, which is scaled to the sensitivity of the utility function, is added to the utility scores~\cite{mcsherry2007}. The sensitivity bounds the difference between the probability of selecting $q$ on any two neighbouring histograms.   

\begin{definition}
The sensitivity of a utility function $u:\mathcal{H} \times Q \rightarrow \mathbb{R}$ to score each query $q \in Q$, denoted by $\Delta u \in[0, 1]$, is $\Delta u = \max_{q \in Q} \max_{H',H'^* \in nbr(H')} ||u(H', q) - u(H'^*,q)||_{1}$.
\end{definition}

\begin{definition}
Given a histogram $H \subseteq \mathcal{H}$, a query set $Q$ and a utility function $u:\mathcal{H} \times Q \rightarrow \mathbb{R}$, the exponential mechanism $EM(H, Q, u)$ outputs a query $q \in Q$ proportional to $exp(\frac{\epsilon u(H,q)}{2\Delta u})$.
\end{definition}

By bounding the sensitivity ($\Delta q$ in Laplace mechanism or $\Delta u$ inExponential mechanism), the privacy budget $\epsilon$ identifies the scale of injected noise. 


\section{Step 1: Private Partitioning} \label{sec:Private Partitioning}
The first step of \textit{DQAM} partitions space into uniform regions $P$ and computes their densities $B$. This stage is independent of the query set and finds $P$ and $B$ as two parameters describing data properties. These parameters are later used in the synthesizing process (step two) to estimate $H'$ as close as possible to $H$ without violating $\epsilon$-differential privacy. $\epsilon_{1}$ is used for computing the cost of a partition and $\epsilon_{2}$ for injecting Laplace noise to the partitions densities, where $\epsilon_{1}+\epsilon_{2} = \epsilon/2$. Algorithm~\ref{alg:partitioning} describes our private partitioning and is discussed below.

\subsection{Quadtree search}
\label{subsec:Quadtree search}
As discussed earlier, the goal is to find partitions in the histogram that are \textit{disjoint} and \textit{uniform}. These requirements are needed to minimize the error in the update step (see s2 in Algorithm~\ref{alg:data and query adaptive estimation}) where we scale all counts in a region with the same magnitude. To achieve this goal, we propose an approach which is a combination of the two approaches described earlier in Section~\ref{sec:mechanism overview}. In particular, instead of generating all possible partitions in the histogram, we approximate them with all partitions that could be generated if the histogram was decomposed using a quadtree. Whilst it might come at the expense of a small cost of utility, it leads to a highly efficient algorithm: (1) Given the partitions in different levels of a quadtree, we can compute the uniformity cost of each partition and add noise to ensure differential privacy; (2) Next, a search procedure expands a quadtree structure can efficiently evaluate all possible combinations and find the least cost set of partitions that are disjoint and cover the entire space of histogram.

\begin{algorithm}[t!]
\begin{tabularx}{\textwidth}{l c l}
Let & $P$ & be the list of partitions, \\ 
    & $B$ & be the corresponding densities.\\
$P = [], B = []$
\end{tabularx}
\caption{Private partitioning using a quadtree.}\label{alg:partitioning}
\begin{algorithmic}
\Procedure{Partition}{$H, P, B,\epsilon_{1}, \epsilon_{2}$}
    \If{$|H| = 1$}
        \State $P.add(H)$
        \State $B.add(density(H) + \nu')$ \Comment{$\nu' \sim Lap(\frac{1}{\epsilon_{2}})$}
    \Else
        \State $pcost \gets cost(H) + \nu$  \Comment{$\nu \sim Lap(\frac{\Delta cost}{\epsilon_{1}})$}
        \State $children \gets split(H)$
        \State $chcost \gets 1/4 * \sum_{i = 1}^{4} cost(children[i]) + \nu_{i}$\\
        \Comment{$\nu \sim Lap(\frac{\Delta cost}{\epsilon_{1}})$} \\
        \If{$pcost - chcost \leq \delta$} 
            \State $P.add(H)$
            \State $B.add(density(H) + \nu')$ \Comment{$\nu' \sim Lap(\frac{1}{\epsilon_{2}})$}
        \Else
            \For{$child \in children$}
                \State{\textbf{return} Partition(child, P, B, $\epsilon_{1}$, $\epsilon_{2}$)}
            \EndFor
        \EndIf
    \EndIf
    \textbf{return} $(P, B)$ \\
    \textbf{end procedure}
\EndProcedure
\end{algorithmic}
\end{algorithm}

The procedure shown in Algorithm~\ref{alg:partitioning} combines the two stages. It takes $H$ as input, and returns the leaves of a (potentially unbalanced) quadtree $P = \{p_{1}, ..., p_{k}\}$ with their densities $B = \{b_{1}, ..., b_{k}\}$. Using the full histogram $H$ as the root node, the algorithm recursively splits $H$ into $4$ regions with a size of power of two using the \textit{split()} function. The noisy cost of each region in the current level (\textit{pcost}) is compared to the \textit{average} of noisy cost of its four children (\textit{chcost}). If the difference between \textit{chcost} and \textit{pcost} is greater than a threshold $\delta$, the partitioning proceeds to the next level. Otherwise, the node is labeled as ``leaf''. Since computing the costs require interacting with the original histogram, a random value $\nu \sim Lap(\frac{\Delta cost}{\epsilon_{1}})$ is added to each region's cost. The algorithm terminates when all regions are labeled as ``leaf'', or the size of each region is $1$ (one cell). Intuitively, when the node has a good level of uniformity, the children would not significantly improve the uniformity cost. 

\begin{figure}[t!] 
    \centering
    \includegraphics[width=0.6\linewidth]{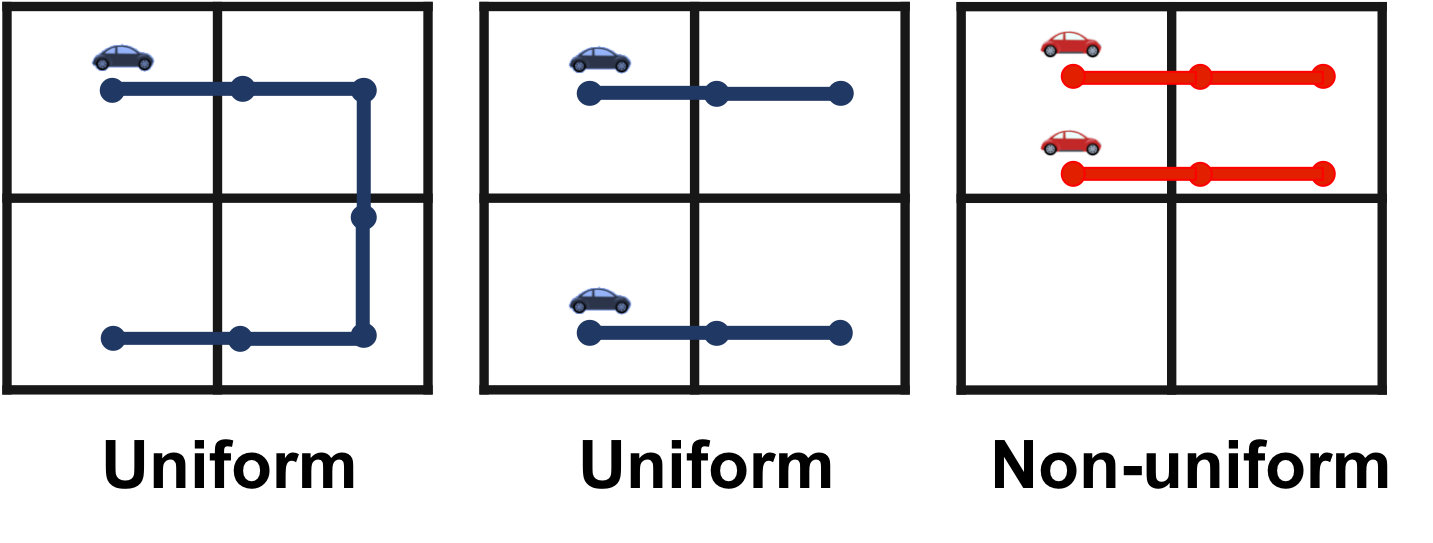}
    \caption{Examples of uniform and non-uniform regions.}
    \label{fig:uniformity-example}
\end{figure}

We define the partitioning cost function to measure the uniformity of a region. Intuitively, a high trajectory density in a part of region incurs a high cost and triggers the algorithm to split the region into smaller ones. Let $j$ denote the cell index in the histogram $H$ and region $p$ be a subset of cell indexes of $H$. We say a region $p$ is uniform when
\[c(H[j])_{j \in p} = \frac{\sum_{i \in p} c(H[i])}{|p|},\]
where $|\cdot|$ and $c(\cdot)$ represent the set size and the cell count, respectively. Fig.~\ref{fig:uniformity-example} shows examples of uniform and non-uniform regions. With this observation, the cost of a region $p$, denoted by \textit{cost(p)}, is the amount that $p$ deviates from being perfectly uniform:
\[cost(p) = \sum_{j \in p} \mid c(H[j]) - \frac{\sum_{i \in p} c(H[i])}{|p|} \mid.\]

\subsection{Computing densities}
After privately identifying uniform regions $P=\{p_{1},...,p_{k}\}$, we need to compute the densities of regions $B=\{b_{1}, ..., b_{k}\}$. The density of region $p_i \in P$, named $b_i$ defines as the number of trajectories in $p_i$ relative to the total number trajectories in the histogram. Given $n$, the total number of trajectories in histogram, the density $b_i$ of region $p_i$ is: \[b_i=\dfrac{\sum_{j \in p_{i}} c(H[j])}{n}.\] 
The function $density(H)$ in Algorithm~\ref{alg:partitioning} denotes computing the density of regions in the histogram $H$.

In the second step of \emph{DQAM}, a region specifies the area to be updated, and its density describes the update magnitude.

\subsection{Formal guarantees}
We analyze Algorithm~\ref{alg:partitioning} for three key aspects: privacy, accuracy and efficiency.

\textbf{Privacy}. We show that step 1 is $\frac{\epsilon}{2}$-differentially private.
\begin{theorem}{Algorithm~\ref{alg:partitioning} satisfies $\frac{\epsilon}{2}$-differential privacy.} 

\begin{proof}
Computing the cost of a region that deviates from a uniform distribution, requires interactions with the histogram, and we must add sufficient noise to ensure differential privacy. We use the Laplace mechanism with $\epsilon_{1}$ to add random noise to the deviation cost. Note that the quadtree is just used as a search procedure to find a least-cost set of regions and we only reveal the leaves of quadtree (the uniform regions) for step 2 and the other regions are removed. Since there is no overlap among the leaves, we can use $\epsilon_{1}$ budget to compute the noisy value of uniformity cost for each region. Hence, the scale of noise is proportional to the sensitivity of regions, denoted by $\Delta cost$, instead of $O(4^{h + 1})$ for the entire quadtree. To ensure privacy, we add a random $\nu_{i} \sim Lap(\Delta cost/\epsilon_{1})$ to the deviation cost of each region. Thus, the total cost of a region is defined in terms of the deviation cost and the error due to the perturbation noise: \[cost(p) = cost(p) + \nu.\]

$\Delta cost$ shows the maximum change in a region's cost when a trajectory is added to (or removed from) the data set. Assuming $k$ as the maximum length of a trajectory, in the worst case, a trajectory increases the count of $k$ cells in a region $p$, and we denote the $k$ cells by set $p'$. We show $\Delta cost \leq 2k$.

Let $cost'(p)$ be the region deviation after adding the trajectory. Assuming that the trajectory only affects the partition $p$ and $|p| \geq k$, we have:
\begin{align}
    cost'(p) & = & \sum_{j \in p'} \mid c(H[j])+1 - \frac{\sum_{j \in p} c(H[j]) + k}{|p|} \mid 
    \\ & + & \sum_{j \in p - p'} \mid c(H[j]) - \frac{\sum_{j \in p} c(H[j]) + k}{|p|} \mid 
    \\ & \leq & k + \sum_{j \in p} \mid c(H[j]) - \frac{\sum_{j \in p} c(H[j]) + k}{|p|} \mid
    \\ & = & 2k + \sum_{j \in p}\mid c(H[j]) - \frac{\sum_{j \in p} c(H[j])}{|p|} \mid
\end{align}
Line $(3)$ is derived using the triangle inequality and line $(4)$ follows from $k > 0$. Using this result we have: \[\Delta cost = cost'(p) - cost(p) \leq 2k.\]

Since a trajectory contributes to every level of the quadtree and thus would be counted multiple times in the overall trajectory count, we divide each trajectory contribution by the number of cells in which it is counted, i.e., its length. Using this approach, we reduce $k$ to $1$, i.e., its actual contribution to the overall trajectory count. This implies that $\Delta cost \leq 2$. We use the technique for computing the cost of regions.

In addition to the regions costs, in Algorithm~\ref{alg:partitioning} we compute the density of regions located in the leaves of the quadtree. Since computing the density requires accessing the original histogram $H$, we use the Laplace mechanism with the remaining privacy budget for step 1, $\epsilon_{2}$, to add $\nu' \sim Lap(\frac{1}{\epsilon_{2}})$ noise to the density count for each region: \[density(p)=\frac{\sum_{j \in p} c(H[j])}{n} + \nu'.\] 
The sensitivity of the density computation is $\leq 1$ as it follows the sensitivity of range queries (see Section~\ref{subsec:DP}). Computing the region costs is $(\frac{\epsilon_{1}}{\Delta cost})$-differentially private and the density calculation is $(\epsilon_{2})$-differentially private. Given $\epsilon_{1} = \epsilon_{2} = \frac{\epsilon}{4}$ and using serial composition for differential privacy, Algorithm~\ref{alg:partitioning} is $(\frac{\epsilon}{4\Delta cost}+ \frac{\epsilon}{4}) \leq \frac{\epsilon}{2}$-differentially private.
\end{proof}
\end{theorem}

\textbf{Accuracy}. The accuracy is measured in terms of the difference between the resultant partitions and the true partitions for the given histogram.

\begin{theorem}{With probability at least $1 - \alpha$, Algorithm~\ref{alg:partitioning} generates partitions with cost at most $TRU + \rho$, where $TRU$ is the cost of true partitions and $\rho=16nlog(n/\alpha)/\epsilon$.}

\begin{proof}
See Appendix~\ref{Appendix A} for the proof. 
\end{proof}
\label{theorem: Accuracy of step 1}
\end{theorem}

\textbf{Efficiency}. Computing optimal partitions in a 2D space are computationally expensive. The complexity of a naive approach is $\Omega(n^{2})$ where $n$ is the number of cells in the histogram. Using the quadtree improves the computational complexity of partitioning to $O(nlog(n))$ as: (1) we partition the data space of a histogram with a fixed cell size and fixed counts, and (2) the structure is compatible with the trajectory data sets and each branch proceeds independently. While the achieved quadtree may not be balanced, the maximum height of the tree is of order $log(n)$ due to (1).

\section{Step 2: Private Synthesizing Process} \label{sec:Private Synthesizing Process}

This section describes the second step of \textit{DQAM}. The inputs are a set of queries $Q$, and the partitions $P=\{p_{1},...,p_{k}\}$ and densities $B=\{b_{1},...,b_{k}\}$ computed for the histogram $H$ in Section~\ref{sec:Private Partitioning}. In this step, we aim to generate a synthetic histogram $H'$ to be as close as possible to $H$. We believe this is the first method proposed that is both data- and query-aware for estimating an optimal spatial histogram $H'$.

\subsection{Synthesizing method}
Our approach is based on the Multiplicative Weights Update Rule (\textit{MWUR})~\cite{hardt2012simple, hardt2010multiplicative}, which is a query-aware method for estimating the distribution of histogram values. Starting from an initial uniform distribution, \textit{MWUR} iteratively selects one query and updates the estimation by rescaling the values based on the query error. Intuitively, if we improve the estimated data for a query with the largest error, the error of other queries overlapping with this query will also be reduced. Hence, instead of considering all the given queries, \textit{MWUR} only considers a small subset of queries to estimate the distribution of data with high utility for all queries. Based on this insight, we create a \textit{data- and query- aware update rule} for privately synthesizing spatial histograms.

As \textit{MWUR} is query-aware but not data-aware, we have to modify \textit{MWUR} when synthesizing spatial histograms. Providing a data-aware approach incurs three challenges:
\begin{enumerate}
    \item The cell counts in a region need to be updated/rescaled proportional to the density of that region. However, \textit{MWUR} considers an equal weight/scale for all cells. Instead of assigning actual counts to each histogram cell, \textit{MWUR} uses the relative proportion of counts (referred to as weight) in each cell. 
    \item \textit{MWUR} only rescales the cell counts contributing to the selected queries. This rescaling strategy is problematic when regions partially overlap with the query area, and some of their cells are outside the query area. For such regions, \textit{MWUR} only rescales the cells inside the query area and leaves the cells outside the query area unchanged. Thus, \textit{MWUR} may degrade the utility of such regions especially those with uniform distribution when answering range queries.
    \item The update function in \textit{MWUR} cannot guarantee the \textit{local sequentiality} of cells, which is a key property of spatial histograms. Ignoring this property can lead to an \emph{inconsistent} output and in turn, reduces the utility.
\end{enumerate}

\begin{algorithm}[t]
\caption{Data and query adaptive estimation of $H$.} 
\label{alg:data and query adaptive estimation}
\begin{algorithmic}
\Procedure{Estimator}{$H, Q, P, B, T, \epsilon_{3}, \epsilon_{4}$}\\
Let $n$ be the total number of traces in $H \subseteq \mathcal{H}$.\\
Let $H'_{0}$ be a uniform estimation over $\mathcal{H}$.
    \For{$i \in T$}
        \State s1. Query Selection: Select $q_{i} \in Q$ using Exponential\\ \hspace{20pt} mechanism with budget $\frac{\epsilon_{3}}{T}$ and the score function \[s(q_{i}, H) = |q_{i}(H) - q_{i}(H'_{i})|.\]
        \State s2. Update: Given the error of $q_{i}$ as $werr_{i}$ perturbed by a \\ \hspace{20pt} noise from $Lap(T/\epsilon_{4})$, rescale the count of cell $j$ in region \\ \hspace{20pt} $p \in P$ with density $b_{p} \in B$ as \[H'_{i+1}[j] \propto H'_{i}[j] . exp(q_{i}[j].werr_{i}.b_{p}/2n).\]
        \State s3. Optimal Estimation: If the update \emph{scaled down} the counts \\ \hspace{20pt} and \emph{inconsistency happened}, compute the nearest histogram \\ \hspace{20pt} as $H'$ that ensures the consistency of cell counts.
    \EndFor
    \textbf{return} $H'$\\
\textbf{end procedure}
\EndProcedure 
\end{algorithmic}
\end{algorithm}

In Section~\ref{subsec:Update Function} we address the first and second challenge by proposing a data- and query-aware update function. Given the uniform regions and their densities from step 1, our update function assigns a scaling factor to each region proportional to its density and preserves the uniformity of the updated regions. To address the third challenge, we need to ensure the \emph{consistency} of the updated spatial histogram. \emph{A consistent spatial histogram guarantees that the answer of a range query monotonically increases with its query area when the larger range query includes smaller range queries.} \cite{Ghane2018Publishing} developed an efficient approach to generate consistent output but the resulting histogram may not be optimal. In Section~\ref{subsec:Optimal Estimation} we discuss consistency for spatial histograms and develop a set of requirements to ensure consistency. We propose a highly efficient method to compute the optimal spatial histogram while ensuring consistency. The privacy budgets $\epsilon_{3}$ and $\epsilon_{4}$ are used for this step to ensure differential privacy. Algorithm~\ref{alg:data and query adaptive estimation} shows our data- and query-adaptive solution for estimating the true histogram.

\subsection{Update function} \label{subsec:Update Function}

Algorithm~\ref{alg:data and query adaptive estimation} shows, in each iteration $i \in T$, the Exponential mechanism consumes $\frac{T}{\epsilon_{3}}$ budget to privately select a query $q \in Q$ such that $q$ has worst error on the current estimation $H'_{i}$. The error of selected query $q$ is then used to update the estimation, i.e., \[werr_{i} = (q(H) + \nu_{i}) - q(H'_{i}).\] 
$\nu_{i} \sim Lap(\frac{T}{\epsilon_{4}})$ is injected noise to the true answer of $q$. 

In each iteration, \textit{MWUR} reduces the estimation error by rescaling the weight of cell counts. Each cell count $H'_{i}[j]$ intersecting the query area is multiplied by a scaling factor $s_{i}$ computed as: 
\[s_{i} \propto exp(q_{i}[j].werr_{i}).\]
$q_i[j]$ is $j^{th}$ element value of query $q$ selected in iteration $i$. $q_i[j] = 1$ if the cell with index $j$ is \emph{inside the query area} and $q_i[j] = 0$ otherwise. The scaling factor (i) only rescales the count of \emph{cells inside a query area} and (ii) all the counts are rescaled equally. Such scaling function requires the original histogram $H$ to be uniformly distributed and fails if the densities in different regions have non-uniform distribution (see Section~\ref{subsec:Quadtree search}). 

We propose a scaling function that (i) updates \emph{all cells in a region} if the region intersects the query area and (ii) rescales a cell count proportional to the density of region that the cell belongs to it. For a region $p \in P$ with density $b_{p} \in B$, our update function rescales the cell $p[j]$ with the following factor:
\[s_{i} \propto exp(q_{i}[j].werr_{i}.b_{p}),\]
where $q_{i}[j] = 1$ if the region $p$ intersects $q$ and $q_{i}[j] = 0$ otherwise. Note now the affected area by $q_{i}[j]$ is expanded beyond the query region. The query error $werr_{i}$ determines the magnitude and direction of update with respect to the selected query. $b_{p}$ amplifies the magnitude of update regarding the region density and $q_{i}[j]$ identifies the extent of update which is the union of query area and the regions intersected the query. According to the scaling factor, we define our update rule for iteration $i \in T$ as follows: 
\begin{equation} \label{formula: update rule}
H'_{i+1}[j] \propto H'_{i}[j] . exp(q_{i}[j].werr_{i}.b_{p}/2n),
\end{equation}
where $n$ is the total number of trajectories in the histogram. 

\subsection{Optimal estimation} \label{subsec:Optimal Estimation}

For a spatial histogram, it is key to preserve local sequentiality of cells, which means any change of a cell value cannot be performed independently without taking into account the adjacent cell values. As mentioned before, the edge values in spatial histograms capture the local sequentiality: an increase in a cell also leads to an increase in an adjacent cell if the edge value indicates that the represented trajectory (or more generally object) spans over both cells. Hence, local sequentiality imposes a \emph{constraint} between each pair of adjacent faces $f_i$ and $f_j$ and their incident edge $e_{ij}$, which has to comply with the constraints $e_{ij} \leq f_i$ and $e_{ij} \leq f_j$ (a trip is always started in a cell per construction and might or might not cross an edge). This constraint is a key requirement for \emph{consistency} in a spatial histogram as any change of data must satisfy it. Violating this constraint can even lead to a negative number of trajectories for a range query on a spatial histogram.

\begin{figure}[t!] 
    \centering
    \includegraphics[width=0.7\linewidth]{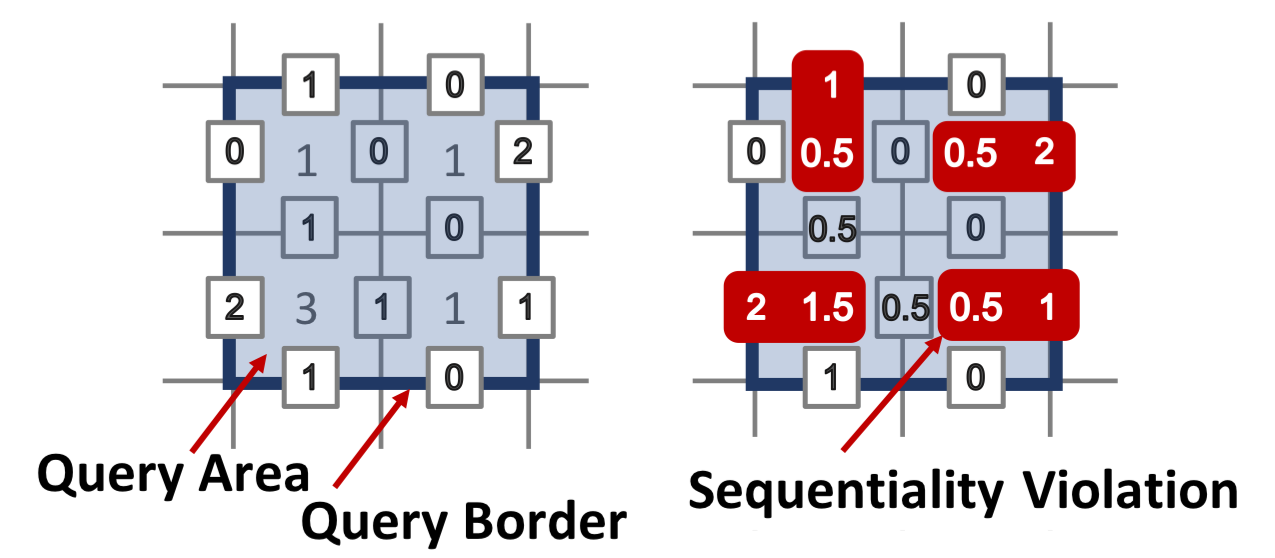}
    \caption{The effective area of a range query and the query borders; the local sequentiality violations caused after updating the edge/face values in the query area by a scaling factor of $0.5$. }
    \label{fig:deo-vio}
\end{figure}

The update rule in Section~\ref{subsec:Update Function} does not consider this constraint when rescaling the face/edge values while computing a synthetic histogram which may lead to inconsistencies. Fig.~\ref{fig:deo-vio} shows an example of the sequentiality violation where the query fully matches to the region. Since the update function rescales the values \emph{inside the query area} it causes violations in edges placing on \emph{borders of the query}. To guarantee the consistency of the synthesized histogram, we describe an approach using linear programming called \emph{consistent inference}. Its aim is to compute the closest consistent histogram to the estimated histogram $H'$ that satisfies the constraints. 
\\
\textbf{Set of Consistency Constraints} ($C$). Every edge count is at most equal to the minimum face count of its adjacent cells:
\begin{equation*}
c(e_{ij}) \leq \min \{c(f_i), c(f_j)\}, \forall e_{ij} \in E'',\textrm{and} \; \forall f_i, f_j \in F''.
\end{equation*}

\begin{definition}{Consistent Inference.}
Let $T$ be the total number of iterations and $H'$ be the estimated histogram in iteration $i \in T$. Given the set of constraints $\mathcal{C}$ between face and edge counts, the \emph{consistent inference} function returns the histogram $H''=\{F'', E''\}$ that satisfies the constraints in $\mathcal{C}$ while minimizing $\|H'' - H'\|$.
\end{definition}

We formulate the consistent inference problem as follows: 

\begin{align*}\label{formula: Lp}
\underset{H''}{\text{minimize}} & \underset{i \in H'}{\mathrm{\sum}} |c(H''[i]) - c(H'[i])| \\
\text{s.t.} & \; c(H''[i]) \geq 0, \forall i \in \{1, ..., |H''|\} \\
& c(e_{ij}) \leq \min \{c(f_i), c(f_j)\}, \forall e_{ij} \in E'',\forall f_i, f_j \in F''.
\end{align*}

We use linear programming to solve this problem. Note that a constraint violation can only happen when the update rule \emph{scales down} the counts or the query error in (\ref{formula: update rule}) is negative. Checking this condition in Algorithm~\ref{alg:data and query adaptive estimation} s3 significantly improves the efficiency of this algorithm.


\subsection{Formal Guarantees} 

We evaluate Algorithm~\ref{alg:data and query adaptive estimation} in three key aspects: privacy, accuracy and efficiency.

\textbf{Privacy}. 
We show Algorithm~\ref{alg:data and query adaptive estimation} is $\frac{\epsilon}{2}$-differentially private.
\begin{theorem}{Algorithm~\ref{alg:data and query adaptive estimation} satisfies $\frac{\epsilon}{2}$-differential privacy.}
\begin{proof}
Given $\epsilon_{3} = \epsilon_{4} = \frac{\epsilon}{4}$, the query selection and computing the error of selected query, each uses $\frac{\epsilon}{4T}$ of the privacy budget. Computing the optimal estimation is a post-processing and hence does not spend any privacy budget. The algorithm makes $T$ calls of the first two stages and according to the serial composition, the budget share accumulates in the calls which result $\frac{\epsilon}{2}$ privacy level for the algorithm. 
\end{proof}
\end{theorem}

\textbf{Accuracy}.
The accuracy is measured in terms of the maximum error bound in answering per $q \in Q$ using the estimated histogram $H'$. The error is because of \textit{perturbation noise} to ensure differential privacy and \textit{estimation error} in rescaling the estimated counts in $H'$ toward the true counts in $H$. 

We show that the accuracy bound of Algorithm~\ref{alg:data and query adaptive estimation} meets the bounds with the state-of-the-art approaches in~\cite{hardt2012simple, Ghane2018Publishing}. However, this is the worst case bound. In our experiments, we show that defining the update rule based on regions and optimizing the estimation with respect to the consistency constraints significantly improves the output accuracy. 

\begin{theorem}{Given a spatial histogram $H$, for any $Q, T \in \mathbb{N}$}, and $\epsilon > 0$, with probability $\geq 1 - 2T / |Q|$, Algorithm~\ref{alg:data and query adaptive estimation} estimates $H'$ such that
\[\max_{q \in Q}{|q(H) - q(H')|} \leq 2n \sqrt{\frac{log|H|}{T}} + \frac{10 T log|Q|}{\epsilon}\]
\begin{proof}
The proof is based on analyzing the effect of perturbation noise and the estimation error on the maximum error in answering queries. See appendix~\ref{Appendix B} for the proof.\vspace{-2mm}
\end{proof} \label{theorem: accuracy}
\end{theorem}

With the fact that $q(H) \in [0, n]$ for all $q \in Q$, we can reduce the bound to less than $n$ by choosing $T$ to be greater than $4log(|H|)$. 

\textbf{Efficiency}. The computational cost of Algorithm~\ref{alg:data and query adaptive estimation} depends on the three steps in each iteration $i \in T$: query selection, updating the estimation, and computing the optimal estimation. For the query set $Q$, computing the score of each query $q \in Q$ has $O(|H|)$ cost where $|H|=r \times c$ is the total number of cells in $H$. Thus, the total cost of query selection is $O(|H||Q|)$. Updating the estimation takes $O(|H|)$ as the cell counts in regions overlapping with the query should be rescaled by the scaling factor. The linear programming solution in the last step has the worst case cost of $O(|H|^{3.5})$~\cite{karmarkar1984new}. The three steps are called for $T$ times that results in the complexity of $O(T(|H||Q| + |H|^{3.5}))$ for the algorithm.

In practice, the total number of queries selected through the iterations (one per iteration $i \in T$), call $Q'$, is much less than the $|Q|$, i.e., $|Q'| \ll |Q|$. The $T$ is also chosen to be a small constant such that $T \cong log(|H|)$ due to the accuracy bound above. These two small constants have a negligible effect on the first part of the complexity cost. In addition, using the violation check after an update can significantly improve the complexity of the algorithm as we do not need to call linear programming function.

\section{Experimental Evaluation} \label{sec:Experiments}

This section evaluates the performance of \textit{DQAM} in different settings explored by recent works including privacy budgets, query sets, data sets, the entire world size (i.e., (1) number of records in the trajectory data set, and (2) number of cells in the histogram), and running time. We contrast the quality of generated histogram with the output of recently proposed algorithms in terms of accuracy and utility. 

\subsection{Experimental setup}
Considering the number of cells (rows times columns) in a histogram as its resolution, we set the resolution of a histogram as a power of $2$. Depending on the speed limit of moving objects, each data set may need a different level of resolution for the histogram. In a data set with $50 km/h$ limit and trajectory sampling rate of 3 GPS coordinates/sec, a histogram with resolution $8$ aggregates the trajectory counts into $2^{8} \times 2^{8}$ cells with $39 m^{2}$ per cell which is fairly accurate for capturing movements in metropolitan areas. Higher resolutions can more accurately capture the movements but are computationally more expensive. Unless otherwise specified, we map the data sets into histograms of resolution $8$. We choose the partitioning threshold $\delta = 4/\epsilon_{1}^{2}$ in Algorithm~\ref{alg:partitioning} to cover the variance of total noise added to the costs of a partition and its potential children. This choice of $\delta$ prioritizes high uniformity in partitioning which leads to highly uniform partitions. We found the computed results stable with this setting for $\delta$. The number of iterations $T$ is chosen from the set $\{10, 12, 14, 16, 18, 20\}$. For each chosen $T$ value, we run the algorithm $5$ times and compute the mean error. Finally, we report the lowest error.

\textbf{Data sets}. One real and one synthetic data set are used in our experiments. The real data set is \textsc{Porto-Taxi} containing the trajectory of taxi-cabs in the city of Porto, Portugal~\cite{moreira2013predicting}. It has about $1.7$ million trajectories of various lengths from $443$ taxi-cabs in one year. \textsc{Porto-Taxi} is a large data set with a highly skewed distribution of trajectories going toward the city center that mapping it to a spatial histogram, results in a histogram with many sparse areas. It is important for the algorithm to be able to capture the distribution information and utilize it by privately estimating the true distribution. The synthetic data set is called \textsc{Melb-Car} which is generated by Minnesota Web-based Traffic Generator (MNTG)~\cite{mokbel2013mntg}. We simulated \num[group-separator={,}]{1000000} vehicle trajectories for $20$ time units in the city of Melbourne, Australia. \textsc{Melb-Car} represents a data set with a uniform distribution where the counts are almost equal in all histogram cells. The two data sets are used to measure the effect of data set distribution on the quality of estimation. To show the effect of data set size $|D|$ on the estimation quality, we use three samples of each data set in the experiments, $|D| \in \{1000, \num[group-separator={,}]{100000}, \num[group-separator={,}]{1000000}\}$. We used a square of $100$km$^2$ as the spatial geometry to map the data set into a histogram. Unless otherwise specified, a data set of $1000$ trajectories from \textsc{Porto-Taxi} is used for the experiment. We note such a small data set with non-uniform distribution results in a sparse histogram and provides the most difficult setting for the algorithm to estimate the true distribution of data. Our results are of high quality even in this hard setting.

\textbf{Queries}. We use queries in the experiments that are uniformly sampled from the domain of histogram $H$. Given $X = [x, x']$ and $Y = [y, y']$ as domains of columns and rows in $H$ respectively, a query is generated as a rectangle $([x_1, x_2], [y_1, y_2])$ where $x_1, x_2 \in X$ and $x_1 \leq x_2$; $y_1, y_2 \in Y$ and $y_1 \leq y_2$. Unless otherwise specified, the default size $|Q| = \num[group-separator={,}]{16000}$ is used for the query set.

\begin{table*}[t!]
    \centering
    \begin{tabular}{|l|c|c|c|c|}
        \hline
        Mechanism & Data-aware & Query-aware & Greedily Consistent & Optimally Consistent \\
        \hline
        \textit{LM} & & \checkmark & &  \\
        \hline
        \textit{MWEM} & & \checkmark & & \\
        \hline
        \textit{DAWA} & \checkmark & \checkmark & & \\
        \hline
        \textit{PriSH} & & \checkmark & \checkmark & \\
        \hline
        \textit{DQAM} & \checkmark & \checkmark & \checkmark & \checkmark \\
        \hline
    \end{tabular}
    \caption{\textit{DQAM} versus existing works in terms of technical and output features.}
    \label{table:comparision}
\end{table*}

\textbf{Measures}. We evaluate the accuracy of estimation in terms of the error in answering queries and measure it using the average $L_{1}$ error per query. In addition, we expect the estimated histogram have high utility in answering arbitrary queries from the class of range queries. The utility of estimated histogram $H'$ is evaluated as the difference between its distribution $H'/n$ and the distribution of original histogram $H/n$:\[RE(H || H') = \sum_{j \in H} c(H[j]) log(\frac{c(H[j])}{c(H'[j])})/n,\] where $n$ is the total number of trajectories in the given data set. This measure is also known as Kullback-Leibler Divergence (KLD) and is used in prior work~\cite{hardt2012simple, Ghane2018Publishing} to evaluate the utility of estimation. This enables us to compare our approach to previous work.

\begin{figure}
    \centering
    \begin{tikzpicture}
        \node[yshift=0cm] (img2)  {\includegraphics[width=0.9\linewidth]{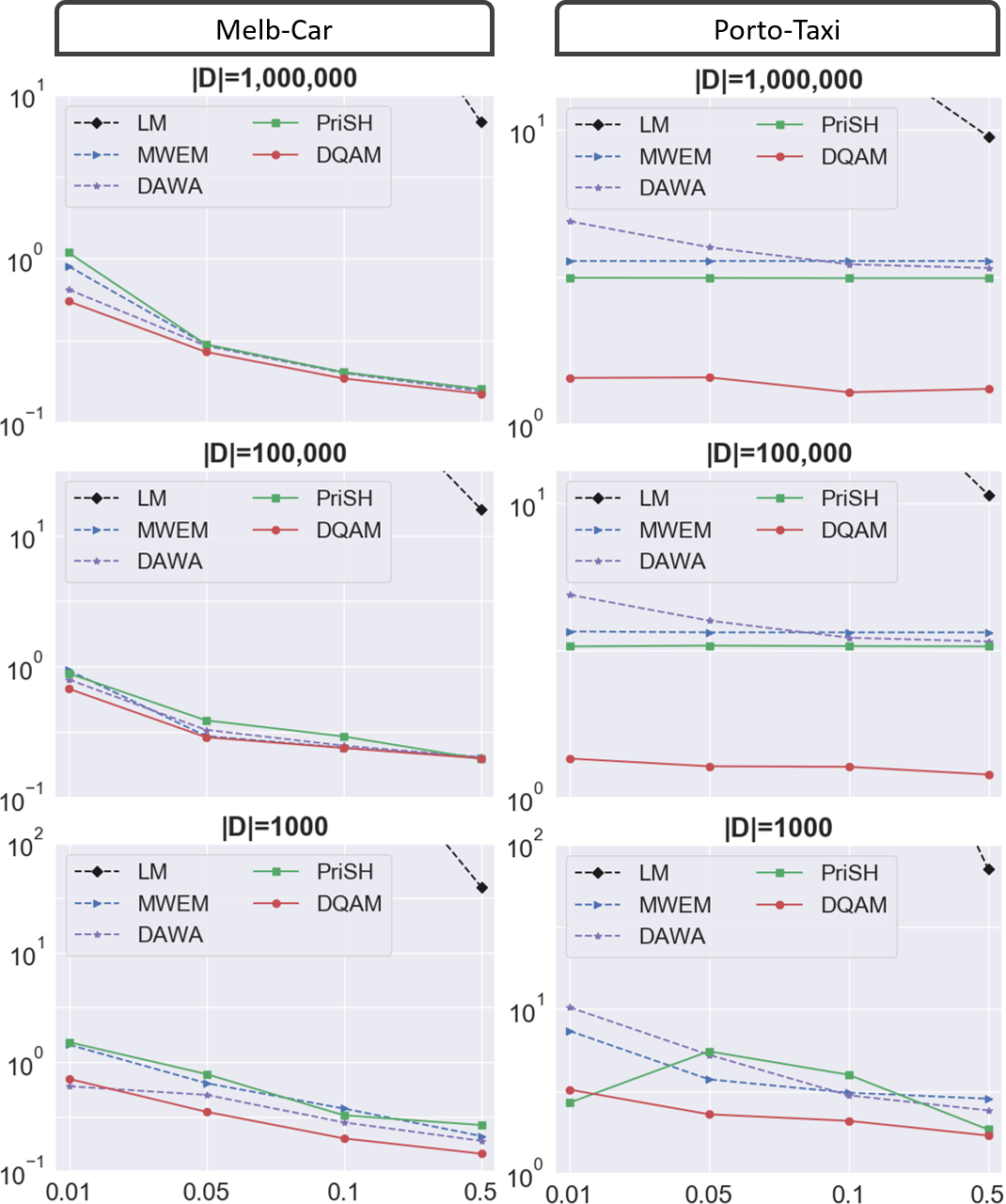}};
        \node[below=of img2, node distance=0cm, yshift=1cm,font=\small] {Privacy Budget $\epsilon$};
        \node[left=of img2, node distance=0cm, rotate=90, anchor=center,yshift=-0.9cm,font=\small] {Average Error (log scale)};
    \end{tikzpicture}
    \caption{The effect of data distribution and size on the error. Dashed line means the output may not have consistency.}
    \label{fig:acc-dsize-dist}
\end{figure}

\textbf{Baselines}. We compare the performance of \textit{DQAM} with four existing works: Laplace mechanism (\textit{LM})~\cite{Dwork2006} as a naive approach, \textit{MWEM}~\cite{hardt2012simple}, \textit{DAWA}~\cite{li2014data} and \textit{PriSH}~\cite{Ghane2018Publishing}. Table~\ref{table:comparision} contrasts the baselines and \textit{DQAM} in terms of technical features, i.e., (1) utilizing data properties (data-aware), (2) utilizing query properties (query-aware), and and the output features, i.e., (3) preserving consistency or local sequentiality (greedily consistent) as well as (4) Optimizing the output while ensuring consistency (optimally consistent). As the table shows, \textit{LM} and \textit{MWEM} only consider the query properties to identify the scale of noise. \textit{LM} is a naive approach in which the noise is directly scaled based on the dependency of queries. In trajectory data, the sensitivity is equivalent to the number of queries which is considerably large. \textit{MWEM}, however, selects a few numbers of queries which significantly reduces the scale of added noise. \textit{DAWA}, is data-aware and query-aware similar to \textit{DQAM} but does not consider the local sequentiality in trajectories (see Section~\ref{sec:Related Works} for more details). Hence, \textit{DAWA} fails to preserve consistency in the output. On the other hand, \textit{PriSH} maintains the consistency of data. However, it only takes advantage of query properties in scaling the noise and generating the output histogram. Additionally, \textit{Prish} fails to generate the optimal histogram with respect to the consistency requirements (see Section~\ref{sec:Related Works} for more details). \textit{DQAM} is the only mechanism that is data- and query-aware, and generates a histogram that is optimal with respect to consistency preservation. 

Since \textit{MWEM} and \textit{DAWA} are designed for 1D histograms with no local sequentiality, we only use the face component to evaluate the algorithms. For these algorithms, we convert the 2D histogram to 1D using Hilbert curves. The number of iterations in \textit{PriSH} and \textit{MWEM} are chosen as explained for \textit{DQAM}, and $\epsilon$ is split evenly between the query selection and the update function in the algorithms. According to~\cite{li2014data}, we consider the ratio $r = 0.5$ for dividing privacy budget $\epsilon$ between the data-aware and query-aware steps in \textit{DAWA}. In our evaluations, we do not report results of \textit{DQAM} and \textit{PriSH} algorithms without consistency. Note that \textit{LM}, \textit{MWEM} and \textit{DAWA} do not guarantee the consistency of generated histogram. Since we only consider the face component to evaluate these mechanisms no penalty is applied for the inconsistency of output. The corresponding results of \textit{LM}, \textit{MWEM} and \textit{DAWA} are depicted by dashed lines in graphs to indicate this limitation.

\subsection{Accuracy Evaluations}
\textbf{Data size and distribution}. Fig.~\ref{fig:acc-dsize-dist} evaluates the accuracy of \textit{DQAM} versus existing works in different data set sizes where results are depicted in log scale. For larger datasets, the counts in the corresponding histogram are larger and hence, less sensitive to noise. That is the algorithms usually have higher accuracy for larger data sets. As the results show, \textit{DQAM} achieves mostly the lowest error specifically for smaller $\epsilon$s. This advantage is more significant on \textsc{Porto-Taxi} data set where \textit{DQAM} accurately captures the sparse/dense regions and estimates the counts properly. In contrast, \textit{PriSH} has an unstable accuracy across different $\epsilon$ values as the heuristic strategy in \textit{PriSH} fails to efficiently estimate histogram in the presence of noise.\textit{MWEM} and \textit{DAWA} work good on uniform distribution in \textsc{Melb-Car}, but their accuracy considerably decreases on non-uniform distribution. The generated histogram, though, is not valid because of inconsistency in data. The error of generated histogram by \textit{LM} is always considerably higher than other mechanisms due to naively scaling the added noise.
Table~\ref{table:error_ratios} reports the standard deviation of accuracy for algorithms for different $\epsilon$ values. As expected, the standard deviation of \textit{LM} is always high, and this naive approach does not show a stable behaviour. \textit{DQAM} has the least standard deviation. In contrast with \textit{PriSH}, \textit{DQAM} employs the optimal consistent estimation resulting in smaller and more stable error which decreases for larger $\epsilon$s. Inconsistency of generated histograms by \textit{MWEM} and \textit{DAWA} result in large variations in the error.

\begin{figure}
\centering
\begin{tikzpicture}
 \node[yshift=0cm] (img2)  {\includegraphics[width=0.9\linewidth]{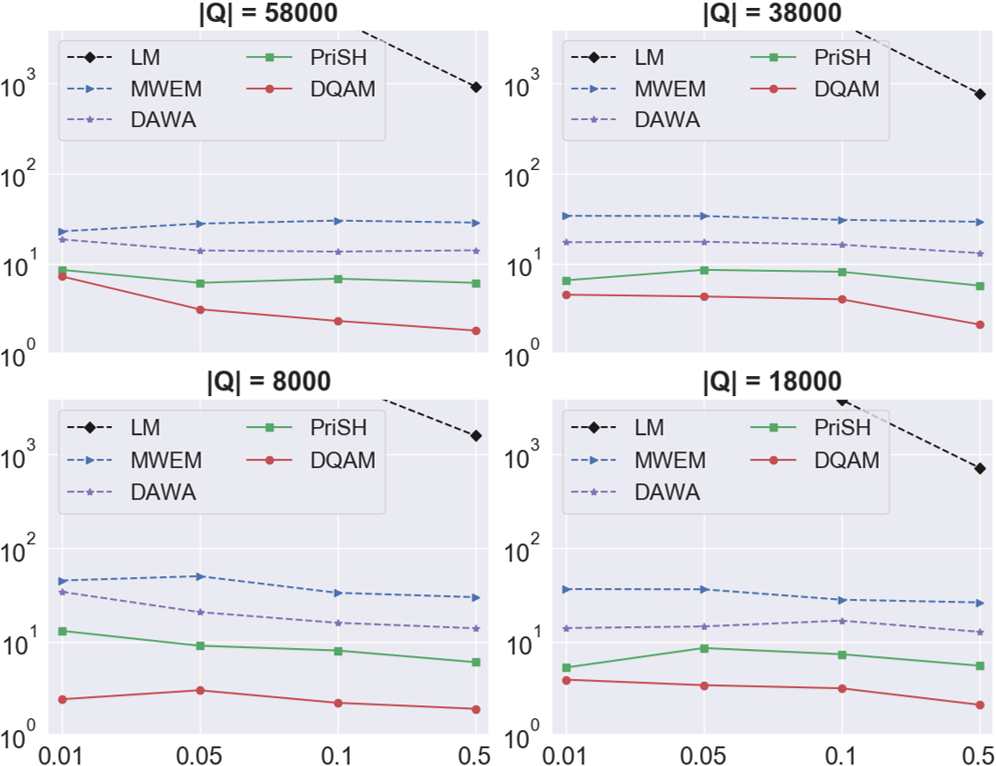}};
 \node[below=of img2, node distance=0cm, yshift=1cm,font=\small] {Privacy Budget $\epsilon$};
 \node[left=of img2, node distance=0cm, rotate=90, anchor=center,yshift=-0.9cm,font=\small] {Average Error (log scale)};
\end{tikzpicture}
\caption{The effect of query set size on the estimation error.}
\label{fig:acc-qsize}
\end{figure}

\begin{table}[]
    \centering
    \begin{tabular}{|c|c|c|c|c|c|}
    \hline
    $\epsilon$ & \textit{LM} & \textit{MWEM} & \textit{DAWA} & \textit{PriSH} & \textit{DQAM} \\
    \hline
    $0.01$ & 500.86 & 5.64 & 3.65 & 2.50  & 1.73  \\
    \hline
    $0.05$ & 500.15 & 4.48 & 2.00 & 2.81 & 1.54 \\
    \hline
    $0.1$  & 429.71 & 3.37 & 1.24 & 1.67 & 0.53 \\
    \hline
    $0.5$  & 285.44 & 3.10 & 0.86 & 0.39 & 0.27 \\
    \hline
    \end{tabular}
    \caption{Standard deviation of \textit{DQAM} error and its competing algorithms for each $\epsilon$ on the sample of \textsc{Porto-Taxi} with $1000$ trajectories.}
    \label{table:error_ratios}
\end{table}

\textbf{Size of query set}. As the set of given queries are randomly generated, the queries may have different shapes. This diversity helps the query selection in \textit{DQAM} to capture the information from the spatial histogram efficiently. We evaluate the effect of changing the query set size on the error of estimated histogram. Fig.~\ref{fig:acc-qsize} depicts the error of \textit{DQAM} versus baseline methods. \textit{DQAM} often achieves higher accuracy comparing to the baselines. Furthermore, the error is almost stable and slightly increases in larger query sets. This behaviour agrees with the Theorem~\ref{theorem: accuracy} in that the size of the query set has a logarithmic effect on the worst case error. 

\begin{figure}
\centering
\begin{tikzpicture}
 \node[yshift=0cm] (img2)  {\includegraphics[width=0.95\linewidth]{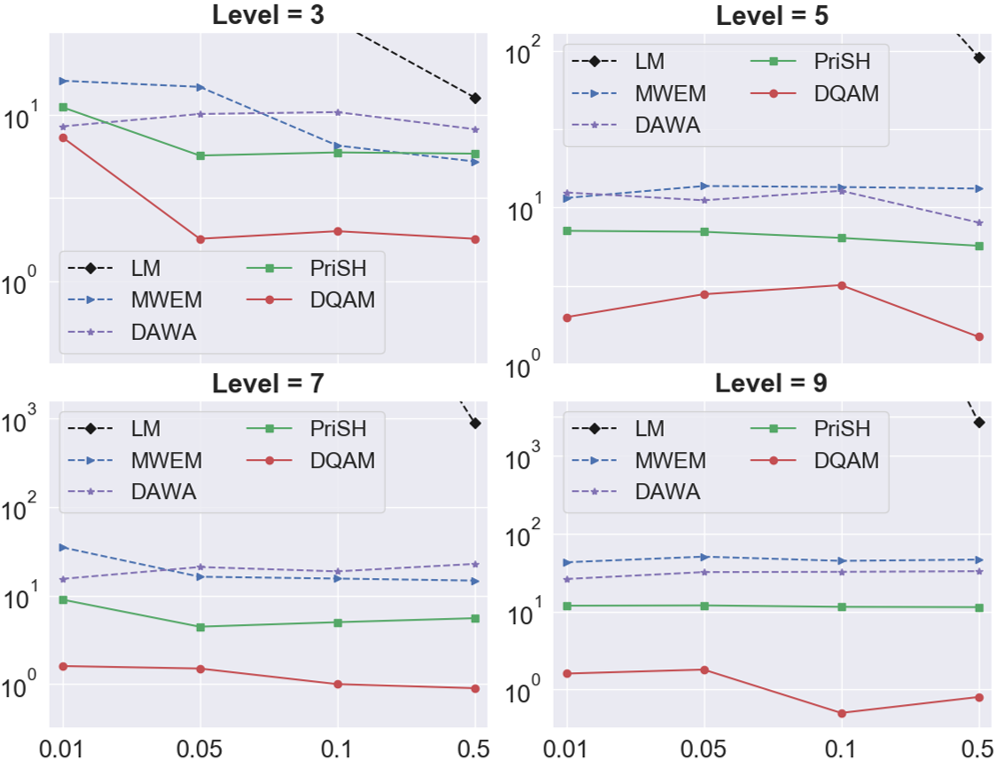}};
 \node[below=of img2, node distance=0cm, yshift=1cm,font=\small] {Privacy Budget $\epsilon$};
 \node[left=of img2, node distance=0cm, rotate=90, anchor=center,yshift=-0.9cm,font=\small] {Average Error (log scale)};
\end{tikzpicture}
\caption{The effect of histogram resolution on the estimation error.}
\label{fig:acc-res}
\end{figure}

\textbf{Resolution of histogram}. Fig.~\ref{fig:acc-res} investigates the effect of changing the histogram resolution on the quality of estimation in \textit{DQAM}. As it was expected, higher resolutions cause higher errors. However, \textit{DQAM} works well in different resolutions comparing to the other algorithms. The significant increment in the error of  \textit{LM}, \textit{MWEM} and  \textit{DAWA} is due to accumulation of error caused by local sequentiality violations.

As mentioned above, the reported results are the average value of $5$ independent runs of an algorithm. The ratio of accuracy improvement achieved by \textit{DQAM} comparing to the competing algorithms changes in different settings. The highest ratio achieved in our experiments was $7.4$ for $|D|=1000$, $|Q|=8000$, resolution $8$ and $\epsilon = 0.01$ comparing the average error of \textit{DQAM} versus \textit{PriSH} shown in Fig.~\ref{fig:acc-qsize}: \textit{PriSH}/\textit{DQAM} $= 14.0654/1.901 = 7.3989$. While the ratio of \textit{DQAM} accuracy versus \textit{LM}, \textit{MWEM} and  \textit{DAWA} is significantly higher (minimum $\geq 60$ for this setting), we did not report it as the output of these algorithms are not consistent due to local sequentiality violations.

\subsection{Utility Evaluation}
\begin{figure}
\centering
\begin{tikzpicture}
 \node[yshift=0cm] (img2)  {\includegraphics[width=0.9\linewidth]{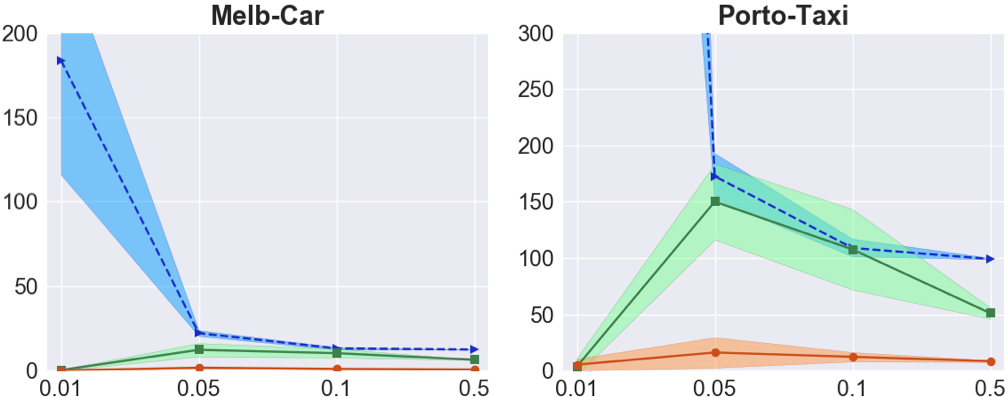}};
 \node[below=of img2, node distance=0cm, yshift=1cm,font=\small] {Privacy Budget $\epsilon$};
 \node[left=of img2, node distance=0cm, rotate=90, anchor=center,yshift=-0.9cm,font=\small] {KLD (lower is better)};
\end{tikzpicture}
\caption{The utility of output by \textit{DQAM} (orange), \textit{PriSH} (green) and \textit{MWEM} (blue) algorithms. Lower KLD means higher utility.}
\label{fig:kld}
\end{figure}

The utility of estimated histogram shows its quality in representing the original histogram. Utility is in fact a measure of quality of generated histogram in answering arbitrary range queries, the ones not included in the query set or not selected in the mechanism process. As mentioned before, We measure the utility of estimated histograms by $KLD$ which is used in prior works. We use this measure to show employing both data and query information in \textit{DQAM} has significant effect in improving the algorithm in detecting sparse/dense regions. Fig.~\ref{fig:kld} depicts that \textit{DQAM} perfectly captures the data distribution properties in both uniform and non-uniform distributions. For clarity in presentation, we focus on \textit{DQAM}, \textit{PriSH} and \textit{MWEM} in this experiment.

\subsection{Efficiency Evaluation}
\begin{figure}
\centering
\begin{tikzpicture}
 \node[yshift=0cm] (img2)  {\includegraphics[width=0.6\linewidth]{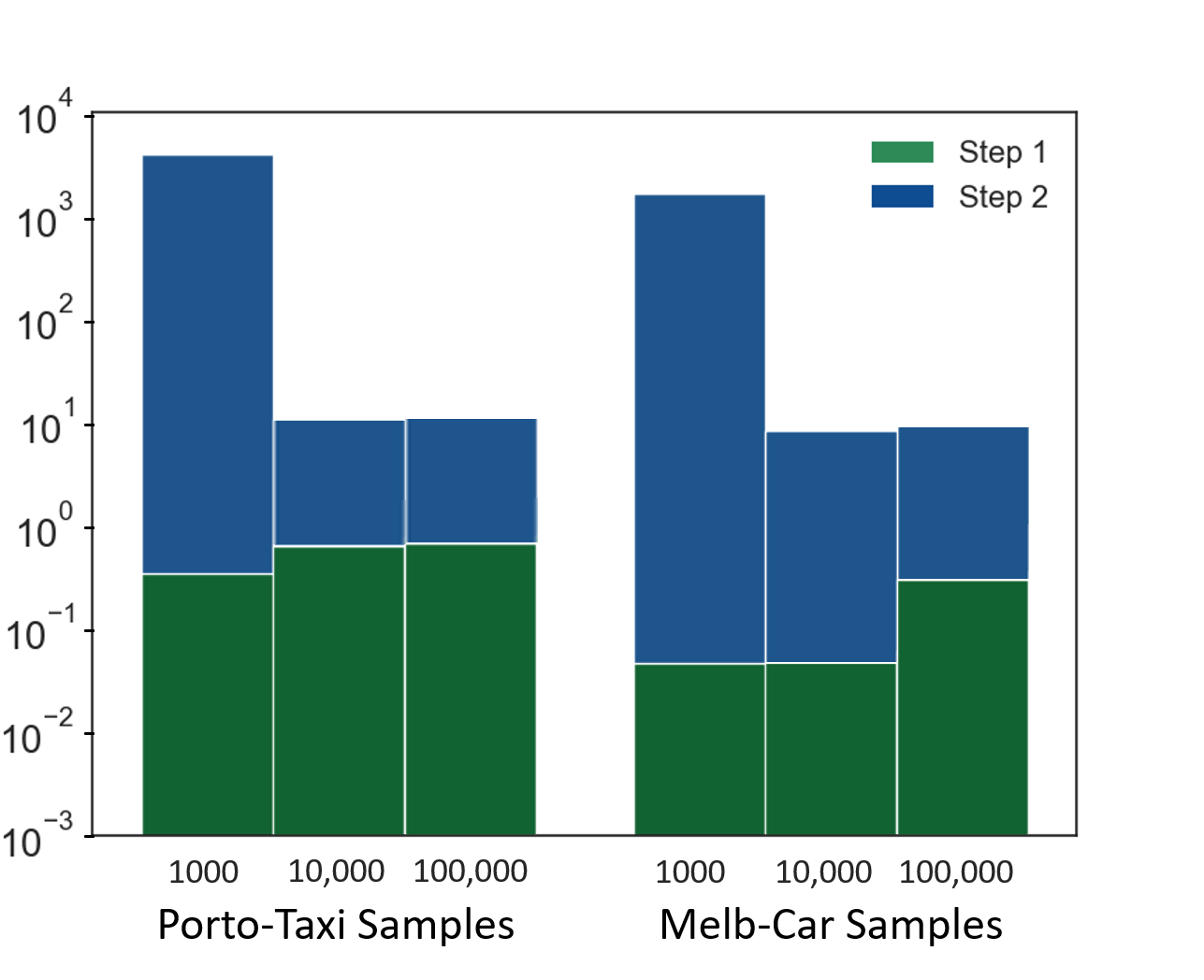}};
 \node[left=of img2, node distance=0cm, rotate=90, anchor=center,yshift=-0.9cm,font=\small] {Time in Seconds (log scale)};
\end{tikzpicture}
\caption{Running time of \emph{DQAM} on small to large datasets.}
\label{fig:running_time}
\end{figure}

Fig.~\ref{fig:running_time} evaluates the running time of \textit{DQAM} on small to large samples of \textsc{Porto-Taxi} and \textsc{Melb-Car} datasets with $\epsilon = 0.1$ and a fixed resolution of histogram. The partitioning (step 1 in \textit{DQAM}) takes less than a second in all settings, and the synthesizing process (step 2 in \textit{DQAM}) mostly contributes to the running time. The total running time for large samples of \textsc{Porto-Taxi} and \textsc{Melb-Car} are $\approx 10$ and $\approx 8$ seconds, respectively. In small samples where the histogram is very sparse, some updates in the synthesizing process cause consistency violations which, in turn, cause the second step to take more time to compute an optimal histogram satisfying consistency.  Note that a violation should be corrected as soon as it happened in the synthesizing process as it can affect the updates in next iterations. Indeed, the accumulation of consistency violations leads the solution to deviate in a wrong direction that it may be far from the optimal point that causes the mechanism to take a long time to compute the optimal estimation in the feasible space of the problem. The total running time for small datasets is $\approx 60$ minutes for \textsc{Porto-Taxi} and $\approx 25$ minutes for \textsc{Melb-Car}. According to the results our mechanism is eminently practical and scalable.

\section{Related Works} \label{sec:Related Works}

Studies on human movement patterns~\cite{de2013unique, xu2017trajectory} show the trajectories are highly identifiable and that coarsening or aggregation barely reduce their uniqueness. Montjoye et al.~\cite{de2013unique} showed that even after coarsening trajectories, $95$\% of individuals could be uniquely identified with only $4$ location points from their trajectories. Recently, Xu et al.~\cite{xu2017trajectory} studied the effectiveness of aggregation in anonymizing trajectories in large-scale data sets and $91$\% of trajectories could be uniquely reconstructed. Hence, private mechanisms with a strong privacy guarantee are needed to publish trajectories.

To address this problem, data-aware mechanisms introduce $\epsilon$-differential privacy to the aggregation techniques~\cite{chen2012differentially, he2015dpt} and reconstruct the trajectory data set from the aggregated data. Even though the mechanisms capture the local sequentiality in trajectories, they are not ideal for efficiently answering range queries. Chen et al.~\cite{chen2012differentially} use a Markov process to model local sequentiality and construct a hierarchical structure such as a prefix tree by grouping the trajectories with the same prefix. Subsequent works~\cite{bonomi2013two, he2015dpt} use most frequent substrings to get higher counts in the leaves and to achieve a better utility. However, a recent work~\cite{naghi2016utility} showed that the mechanisms fail to preserve trajectory properties such as their length, and thus cannot guarantee the accurate distribution of trajectories. Even the use of a hierarchical structure, i.e., the constructed noisy prefix tree, is computationally expensive for range queries. Given a prefix tree of height $l>0$ with a branching factor $a>1$, answering a range query has a computational cost of $O(a^{l+1})$. Our mechanism, however, publishes a 2D histogram that maintains the local sequentiality and costs $O(k)$ on range queries where $k$ is the size of query area.

Monreale et al.~\cite{monreale2013privacy} define the dependency between cells instead of points by mapping the trajectories to a grid to count the movement frequencies between the adjacent cells. However, a frequency vector only maintains the number of transitions for a group of observations without information about the spatial adjacency of two vectors. This information is crucial for a range query as it counts the vectors overlapping the query area. In our work, we use a spatial histogram that is a grid but captures both the cell counts and the spatial adjacency of trajectories. 

Related work in coarsening the data space are in two categories: data-aware mechanisms~\cite{cormode2012differentially, qardaji2013differentially} and query-aware mechanisms~\cite{Dwork2006, hardt2012simple,li2014data}. The data-aware mechanisms use some data properties to adaptively partition the data space and optimize the error of published data in answering range queries. However, the models are developed for location points rather than trajectories. Cormode et al.~\cite{cormode2012differentially} build a hybrid tree in which a kd-tree partitions the space into coarse density regions and a quadtree maps the dense areas into finer resolutions. Qardaji et al.~\cite{qardaji2013differentially}, generate partitions to get uniform regions of data. The partitioning condition is based on the density of location points in regions. The data-aware partitioning techniques make use of information about the data distribution to enhance accuracy. However, the mechanisms are not able to capture local sequentiality of location points which is crucial in representing trajectories. The query-aware mechanisms, on the other hand, take a set of range queries to get information about the data distribution. Laplace mechanism~\cite{Dwork2006} is a naive query-aware model that adds an independent noise to the answer of each query in the query set. Hardt et al.~\cite{hardt2012simple} developed \textit{MWEM} that optimizes the accuracy of published data for the class of range queries. By laying a grid with a fixed cell size on the data space, \textit{MWEM} iteratively learns the density of cells using the noisy answer of queries. Li et al.~\cite{li2014data} proposed a data- and query-aware (\textit{DAWA}) which employs data properties as well as query properties in publishing data under differential privacy. A comparative analysis of space decomposition mechanisms by Hay et al.~\cite{hay2016principled} showed \textit{MWEM} and \textit{DAWA} achieve high accuracy in answering range queries on location data, not trajectories. Since the mechanisms are not able to capture the local dependency between locations which is crucial in studying trajectories. In this paper, our proposed mechanism uses spatial histograms in which the edge component maintains the local sequentiality and the learning process ensures preserving this property in the published spatial histogram.

Recently, Ghane et al.~\cite{Ghane2018Publishing} developed a mechanism named Private Spatial Histogram \textit{PriSH} for range queries on trajectories. \textit{PriSH} publishes a synthetic spatial histogram under $\epsilon$-differential privacy. It is a query-aware mechanism that extends the idea of \textit{MWEM} in~\cite{hardt2012simple}. \textit{PriSH}, takes a spatial histogram and a query set as input and utilizes the correlation between the queries to estimate the distribution of the original histogram privately. To maintain the consistency in the histogram, \textit{PriSH} uses a \emph{heuristic} approach that may result in a histogram far from the original spatial histogram. The reason lies in the \emph{heuristic} approach that locally ensures consistency and may lead to overcorrection. In this paper, we propose a data- and query- aware mechanism that utilizes the trajectories density in different regions as well as the given queries correlation to estimate the optimal spatial histogram with significantly higher utility. Our query-aware strategy employs a linear programming approach to provide a guarantee on optimally consistent histogram which leads to significant improvement in the utility of results.

\section{Conclusion} \label{sec:Conclusion}
We proposed \textit{DQAM}, a data and query-aware mechanism for privately publishing spatial histograms and answering range queries on trajectories. The first step of \textit{DQAM} identifies disjoint uniform regions in a histogram and computes the density of each histogram region. It uses a quadtree as it enables efficient space partitioning. Based on the quadtree, we can measure the uniformity for each step of the partitioning. The partitions and densities are then used in the second step of \textit{DQAM} for learning the true distribution of histogram. The learning process uses a given query set as well as the partitions and densities from the first step to estimate the true distribution and to improve the quality of estimation in each iteration. Our experiments show that \textit{DQAM} significantly improves the accuracy over the state of the art methods, specifically on data sets with a non-uniform distribution while ensuring consistency, i.e., local sequentiality. Our experimental results show that we have improved utility by a factor 7.4 (see Fig.~\ref{fig:acc-qsize}). In the future, we will focus on differentially private trajectory representations that retain high data utility and thus can be used for any spatial query type as current techniques cannot achieve the required data utility.

\bibliographystyle{plain}
\bibliography{main}

\appendices

\section{Proof of Theorem~\ref{theorem: Accuracy of step 1}}
\label{Appendix A}
Let $TRU$ be the cost of true partitions and $L_{\rho}$ be the set of all possible partitions that for $\textbf{P} \in L_{\rho}$ the total cost of partitioning is $cost(\textbf{P}) \leq TRU + \rho$. $\tilde{L_{\alpha}}$ represents the complement of set $L_{\alpha}$. Let $\nu$ be the Laplace random noise with scale $b = \Delta cost / \epsilon_{1}$ added to each region $p \in \textbf{P}$. Given $|H|=n$ as the number of cells in the histogram, if $\norm{\nu} < \frac{\rho}{2n}$, the total noise added to the partition cost is $n.\frac{\rho}{2n} = \frac{\rho}{2}$ because a partitioning can have at most $n$ regions. This means with such noise still no partitioning in $\tilde{L_{\alpha}}$ will be generated. Given Algorithm~\ref{alg:partitioning} as $\mathcal{A}$, we have:
\begin{align}
P(\mathcal{A} \in L_{\rho}) & \geq & P(\nu < \frac{\rho}{2n} \forall p \in \textbf{P}) \\
& = & 1 - P(\exists p \in \textbf{P}, that, \nu \geq \frac{\rho}{2n})\\
& \geq & 1 - |\textbf{P}|P(\nu \geq \frac{\rho}{2n})  \\
& = & 1 - |\textbf{P}|exp(-\frac{\rho}{2bn}) \\ & = & 1 - \alpha 
\end{align}
The line $(7)$ is achieved by union bound. The line $(8)$ follows the definition of Exponential mechanism with scale $b$ because the absolute value of a random variable from Laplace distribution has an exponential distribution. Now, we can compute the $\rho \geq 2.b.n.log(n/\alpha) = \frac{4.n.log(n/\alpha)}{\epsilon_{1}} =  \frac{16.n.log(n/\alpha)}{\epsilon}$

\section{Proof of Theorem~\ref{theorem: accuracy}}
\label{Appendix B}

The proof follows the analyze of accuracy in~\cite{hardt2012simple}. We show that our new update rule and later the linear programming do not affect the worst case bound of accuracy. Without further amendments, the proof in~\cite{hardt2012simple} carries over from 1D to 2D for the spatial histogram in our work. It first computes a bound on the perturbation noise regarding the query error. Second, it uses the relative entropy to bounds the estimation error imposed by the update rule. Last, it combines the two bounds to achieve the worst case error bound on the accuracy. Our improvements affect the second step of analysis and we show it does not change the bound of estimation error. 

Our update rule rescales the count of each cell $x \in H'$ proportional to the density of region that $x$ belongs to that, i.e., $b_{x} \in B$. After scaling the counts in each round, the optimal estimator multiplies the cell counts by a multiplier $\alpha_{x}$ where $\alpha_{x} \geq 0$. Considering a maximum multiplier in each round, we define $\alpha$ as the maximum bound on the multipliers across all rounds, i.e., \[\alpha = \max_{i \in T}\max_{x \in H'}{\alpha_{x}}.\]

According to the definition of relative entropy, the difference between the $H$ and $H'$ after each update round improves as:\vspace{-4mm}

\begin{eqnarray} \label{formula: entropy}
\psi_{i-1} - \psi_{i}  =  \sum_{x \in H}c(H[x]) log (\frac{c(H'_{i}[x])}{c(H'_{i-1}[x])})/n,
\end{eqnarray}
where the ratio $c(H'_{i}[x]) / c(H'_{i-1}[x])$ can be written as $\frac{\alpha}{\beta_{i}}sp_{i}$ such that $sp_{i} = exp(q_{i}[x]\eta_{i}b_{x})$ with $\eta_{i} = (m_{i} - q_{i}(H'_{i-1})/2n$ and $b_{x} \in B$ is the density of region that the cell $x$ belongs to it. $\beta_{i}$ is the re-normalization factor. As the cells may have different density coefficients, we consider $\textbf{b} = \{\textbf{b} \in B \mid \forall b_{i} \in B, \textbf{b} \geq b_{i}\}$. Expanding the equation gives: \vspace{-4mm}

\begin{eqnarray*}
    \psi_{i-1} - \psi_{i} & = & log(\alpha) + \textbf{b}\frac{\eta_{i}}{n}q(H) - log(\beta_{i}).
\end{eqnarray*}

Using the Taylor expansion of $sp_{i}$ and $q_{i}[x]^{2}\leq 1$, the $\beta_{i}$ bound will be: \vspace{-4mm}

\begin{eqnarray*}
    \beta_{i} & = & \sum_{x \in H'}{(\alpha . sp_{i} . c(H'_{i-1}[x])) / n} 
    \\& \leq & \sum_{x \in H'}{\frac{\alpha}{n}(1 + \textbf{b}q_{i}[x])\eta_{i} + \textbf{b}^{2}\eta_{i}^{2})c(H'_{i-1}[x])}
    \\ & = & \alpha (1 + \frac{\textbf{b}\eta_{i}}{n}q_{i}(H'_{i-1}) + \textbf{b}^{2}\eta_{i}^{2}).
\end{eqnarray*}
Inserting this bound to the equality~\ref{formula: entropy} and that $log(1+x) \leq x$ and $\textbf{b} \leq 1$, cancels $log(\alpha)$ and gives:

\begin{eqnarray*}
    \psi_{i-1} - \psi_{i} & \geq & \textbf{b}\frac{\eta_{i}}{n}q(H) - log(1 + \textbf{b}\frac{\eta_{i}}{n}q_{i}(H'_{i-1}) + \textbf{b}^{2}\eta_{i}^{2}) 
    \\ &\geq & \eta_{i}(q_{i}(H) - q_{i}(H'_{i-1})/n - \eta_{i}^{2}.
\end{eqnarray*}
By introducing the definition of $\eta_{i}$ to this result, it gives:

\begin{eqnarray*}
        \psi_{i-1} - \psi_{i} \geq (\frac{q_{i}(H'_{i-1}) - q_{i}(H)}{2n})^{2} - (\frac{m_{i} - q_{i}(H)}{2n})^{2},
\end{eqnarray*}

which is the same bound on the estimation error in~\cite{hardt2012simple}. The remaining of the proof follows the accuracy analysis in~\cite{hardt2012simple}.

\end{document}